\documentclass[]{spie}  

\usepackage{amsmath,amsfonts,amssymb}
\usepackage{graphicx}
\usepackage{subcaption}
\usepackage{float}
\usepackage[colorlinks=true, allcolors=blue]{hyperref}

\title{
Toward a robust lesion detection model in breast DCE-MRI: adapting foundation models to high-risk women}

\author[a, b]{Gabriel A.B. do Nascimento}
\author[a]{Vincent Dong}
\author[a, b]{Guilherme J. Cavalcante}
\author[a]{Alex Nguyen}
\author[b]{Thaís G. do Rêgo}
\author[b]{Yuri Malheiros}
\author[c]{Telmo M. Silva Filho}
\author[a]{Carla R. Zeballos Torrez}
\author[a]{James C. Gee}
\author[a]{Anne Marie McCarthy}
\author[a]{Andrew D. A. Maidment}
\author[a]{Bruno Barufaldi}

\affil[a]{University of Pennsylvania, Philadelphia, PA, US}
\affil[b]{Federal University of Paraíba, João Pessoa, PB, Brazil}
\affil[c]{University of Bristol, Bristol, UK}

\begin{document} 
\maketitle

\begin{abstract}
Accurate lesion detection in breast MRI is essential for early cancer diagnosis, particularly in high-risk populations. In this study, we present a classification pipeline that adapts a pretrained foundation model—Medical Slice Transformer (MST) to perform breast lesion classification using dynamic contrast-enhanced MRI (DCE-MRI). Leveraging DINOv2-based self-supervised pretraining, the MST model provides robust per-slice feature embeddings. We extract these representations and use them to train a Kolmogorov–Arnold Network (KAN) classifier, which offers a more flexible and interpretable alternative to traditional convolutional neural networks. The KAN architecture enables localized nonlinear transformations through adaptive B-spline activations, enhancing the model's ability to distinguish between benign and malignant lesions in imbalanced and heterogeneous clinical data. Our results show that the MST+KAN pipeline outperforms MST transformer based classifier, achieving improved performance metrics (AUC=0.80$\pm$0.02) while maintaining interpretability via attention-based heatmaps. This study highlights the superiority of combining foundation model embeddings with advanced classification strategies to build robust and generalizable breast MRI analysis tools.
\end{abstract}


\section{INTRODUCTION}
\label{sec:intro}  


Breast magnetic resonance imaging (MRI) is a sensitive modality that offers superior visualization of breast parenchyma and lesions compared to mammography or ultrasound \cite{10.1001/jamainternmed.2013.11958, 10.1001/jamainternmed.2013.11963, Lee2021-zz, Killelea2013-fy}. Dynamic contrast-enhanced MRI (DCE-MRI) adds functional and morphological information, improving lesion detectability. However, DCE-MRI is still challenged by high false-positive rates (FP), particularly in BI-RADS category 4 lesions, which has a broad probability of malignancy (2-95\%) \cite{Strigel2017, Raza2008}. Moreover, radiologist interpretations are highly subjective and vary between readers. As a result, this has led to a large number of negative biopsies (70-80\% of biopsied lesions are benign), increasing healthcare costs, patient anxiety, and clinical burdens \cite{Vlahiotis2018}.


Deep learning has advanced lesion detection by capturing complex image features, but  the variability in patients, protocols, and scanners limits generalizability\cite{Yu2022}. Self-supervised foundation models like DINO~\cite{wang2025foundation}, trained on large natural image datasets, offer transferable visual representations without manual and multiple annotations. However, DINO is not optimized for the fine-grained contrast and anatomical detail essential in medical imaging, often resulting in reduced performance and interpretability. Bridging this domain gap requires task-specific adaptation and fine-tuning to align general-purpose vision features with clinical needs, enabling models to extract relevant and interpretable features that are crucial for trust and adoption in healthcare.



In this work, we adapt the Medical Slice Transformer (MST)\cite{Jang2022M3T}, a pre-trained slice-based transformer, to our local patient population. We perform inference with the pre-trained MST to extract general breast MRI feature embeddings that are then used to train a new classifier to detect breast lesions in high-risk women undergoing breast cancer screening with MRI. This approach allows us to effectively leverage pre-trained representations; general imaging and breast cancer features, to adapt the MST model to our specific clinical task and dataset; detecting BI-RADS 4 lesions. Our goal to fine-tuning breast imaging foundation models to develop specialized diagnostic-specific tools for challenging and underrepresented populations.

\section{MATERIALS AND METHODS}
\label{sec:methods}

\subsection{Study Population}
Biopsy-confirmed benign (n = 4,900) and malignant (n = 1,214) breast DCE-MRI studies were collected from a retrospective, IRB-approved and HIPAA-compliant study conducted at the University of Pennsylvania Health System (UPHS) between January 2016 and January 2023. The imaging protocols for each study included T1-weighted axial sequences acquired both pre and post-contrast. We excluded studies malignant studies where cancer laterality was not provided. Studies with prior breast cancers and implant were also excluded.

\subsection{Data Preprocessing}
For analysis, each MRI volume was divided into separate image stacks for the left and right breasts. These volumes were resampled to a resolution of 1.0x1.0x1.0 $\mathrm{mm}^3$ to standardize voxel dimensions. N4 bias field correction\cite{Tustison2010} was used to correct for intensity inhomogeneities in the images and piecewise linear histogram equalization (PLHE)\cite{10.1145/3383812.3383830} was used to normalize the images to reduce scanners variance and acquisition settings.  

Further preprocessing follows the methods described in Müller-Franzes et al. \cite{Muller-Franzes2025}. In brief, subtraction images were generated from pre-contrast images and their corresponding post-contrast images. The images were then cropped and padded to a uniform dimension of 256×256×32 voxels.

\subsection{Model Implementation}
We adapt the Medical Slice Transformer (MST) architecture proposed by Müller-Franzes et al. for our UPHS cohort \cite{Muller-Franzes2025}. The model architecture is a slice-wise Vision Transformer (ViT) that leverages the pre-trained DINOv2 foundation model for powerful feature extraction \cite{oquab2023dinov2}. The DINOv2 model includes a pre-trained image encoder that utilizes the imaging features learned from the ImageNet dataset, which are robust to variations in lighting, texture, and object appearance. The MST model was specifically trained to diagnose breast cancer on breast MRI data using the DUKE-Breast-Cancer-MRI dataset\cite{saha2021dce}.

The MRI data are processed per slice, where each 2D axial slice of size 256×256 is treated as an independent input to the model. Although the model operates at the slice level, the final classification is performed at the patient level, using the aggregated information across all slices to predict whether the lesion is benign or malignant. This approach allows the model to capture localized slice-level features while aligning predictions with patient-level malignancy labels.

\subsection{Transfer Learning Classification}
We explore the effectiveness of transfer learning by fine-tuning the pre-trained MST model to classify breast lesions as benign or malignant using our collected cohort of high-risk women imaged with breast MRI. The MST model, initialized with DINOv2 weights, provides a rich feature representation tailored to breast MRI data, making it particularly well-suited for transfer learning in this binary classification task.

To perform transfer learning, we identify the last linear layer in the MST model that is responsible for the final classification. We freeze this layer and extract the resulting feature vector, which serves as the per-slice ViT embedding of the input image. The feature vectors extracted from our dataset are then used as training data for a new binary classification model.


\textbf{Neural Network Architecture:}
We explore the Kolmogorov-Arnold network (KAN)\cite{liu2024kan} model architecture as a binary classification model trained on the feature embeddings extracted from the MST model. KANs utilize adaptative B-spline interpolators as activation functions, allowing the model to be more flexible and interpretable than traditional multi-layer perceptron (MLP) dense layer approaches.

The KAN architecture consists of an input layer, two hidden layers with 128 and 64 neurons, respectively, and a single output logit for binary classification. To stabilize training and improve generalizability, the model employs a Sigmoid Linear Unit(SiLU) base function combined with adaptive B-spline interpolators that have learnable grid points and trainable spline parameters, enabling localized nonlinear transformations in the feature space.

\textbf{Model Training:}
The model was trained and evaluated using a 5-fold stratified cross-validation with non-overlapping groups to ensure robust performance assessment. To address class imbalance, we randomly sample a volume from the cohort of benign patients such that each sample consists of one breast. Within each fold, we also use Borderline-SMOTE \cite{10.1007/11538059_91} to further account for the class imbalance. The model was trained for 500 epochs with early stopping (patience = 5) and a batch size of 64. We used the AdamW optimizer with a learning rate of 1e-6. Focal loss \cite{lin2018focallossdenseobject} was used as loss function because it places greater emphasis on misclassified samples.

\section{RESULTS AND DISCUSSION}

\begin{figure}[!ht]
    \centering
    \includegraphics[width=17cm]{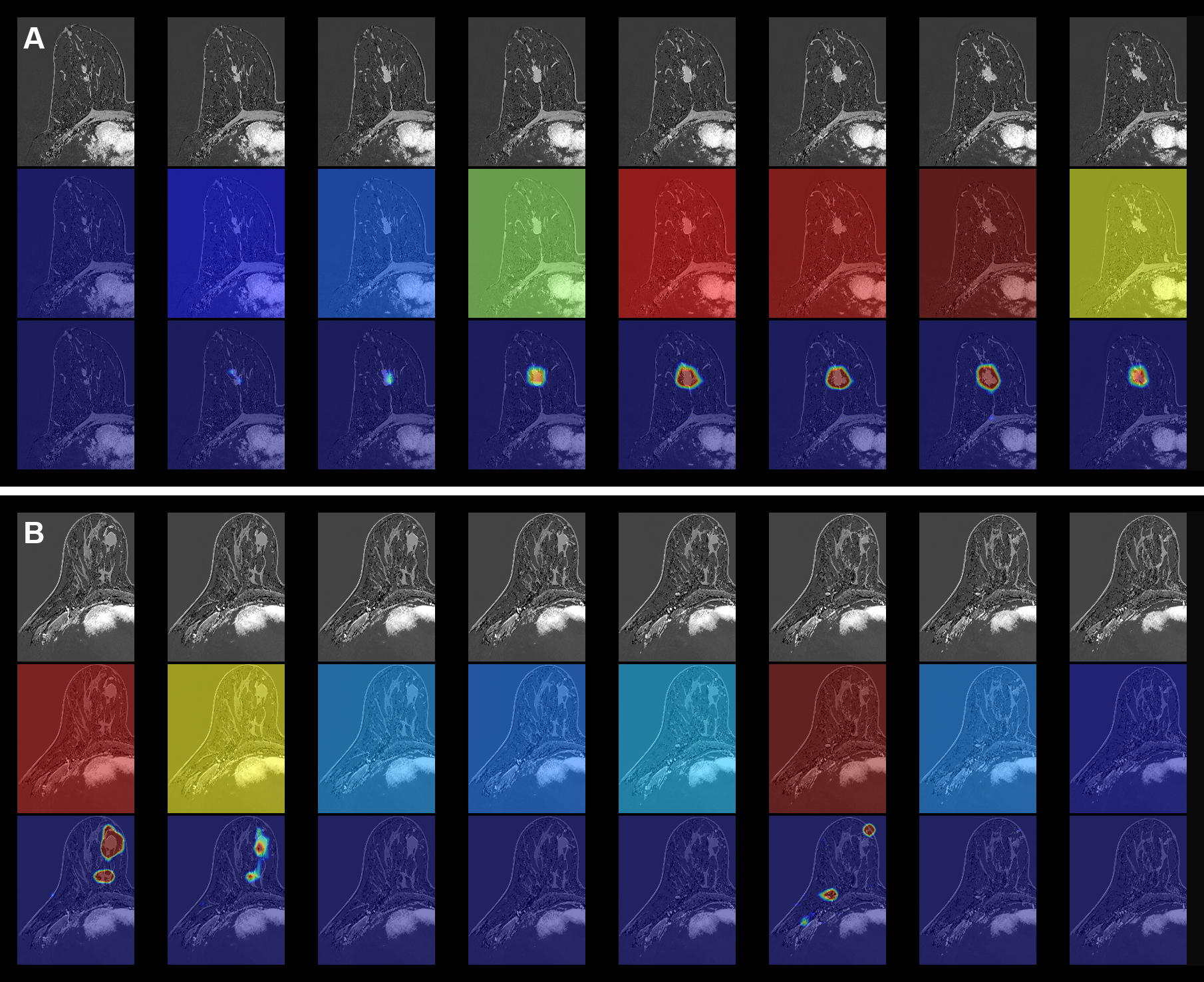}
    \caption{MST model attention for two representative cases. (Top row) Subtraction images from consecutive axial slices. (Middle row) Model attention at the slice-wise level, with cooler colors indicating less attention and warmer colors indicating higher attention. (Bottom row) Model attention identifying potential lesions. Note in sample B, some attention is incorrectly assigned, underscoring limitations in localization accuracy.}
    \label{fig:sub1}
\end{figure}

Figure \ref{fig:sub1} shows attention heatmaps overlaid on subtraction MRI slices, highlighting the regions most influential in the model’s lesion detection decisions. Across different patient cases, the attention consistently concentrates on lesion sites, even in the presence of dense breast tissue or small lesions. In cases with larger lesions, the heatmaps demonstrate broader but still localized focus. The model also maintains attention specificity across varied scanner types, indicating robustness to imaging heterogeneity. These visualizations confirm that the model captures lesion-relevant features to determine malignancy without explicit segmentation supervision.

\begin{figure}[!ht]
    \centering
    \begin{subfigure}[b]{0.48\textwidth}
        \centering
        \includegraphics[height=7.8cm]{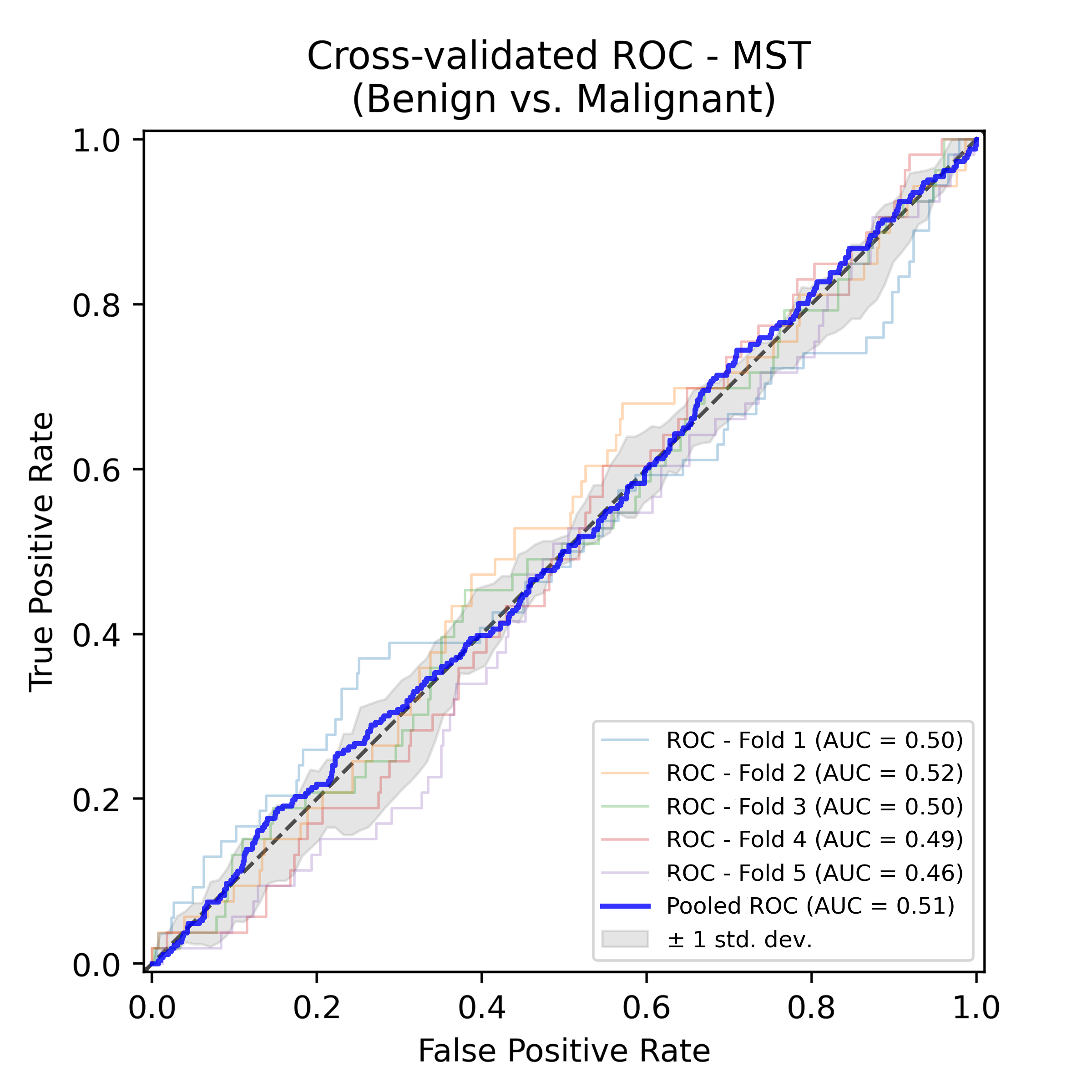}
        \caption{ROC-AUC curve of the MST classifier.}
        \label{fig:roc_mst}
    \end{subfigure}
    \hfill
    \begin{subfigure}[b]{0.48\textwidth}
        \centering
        \includegraphics[height=7.8cm]{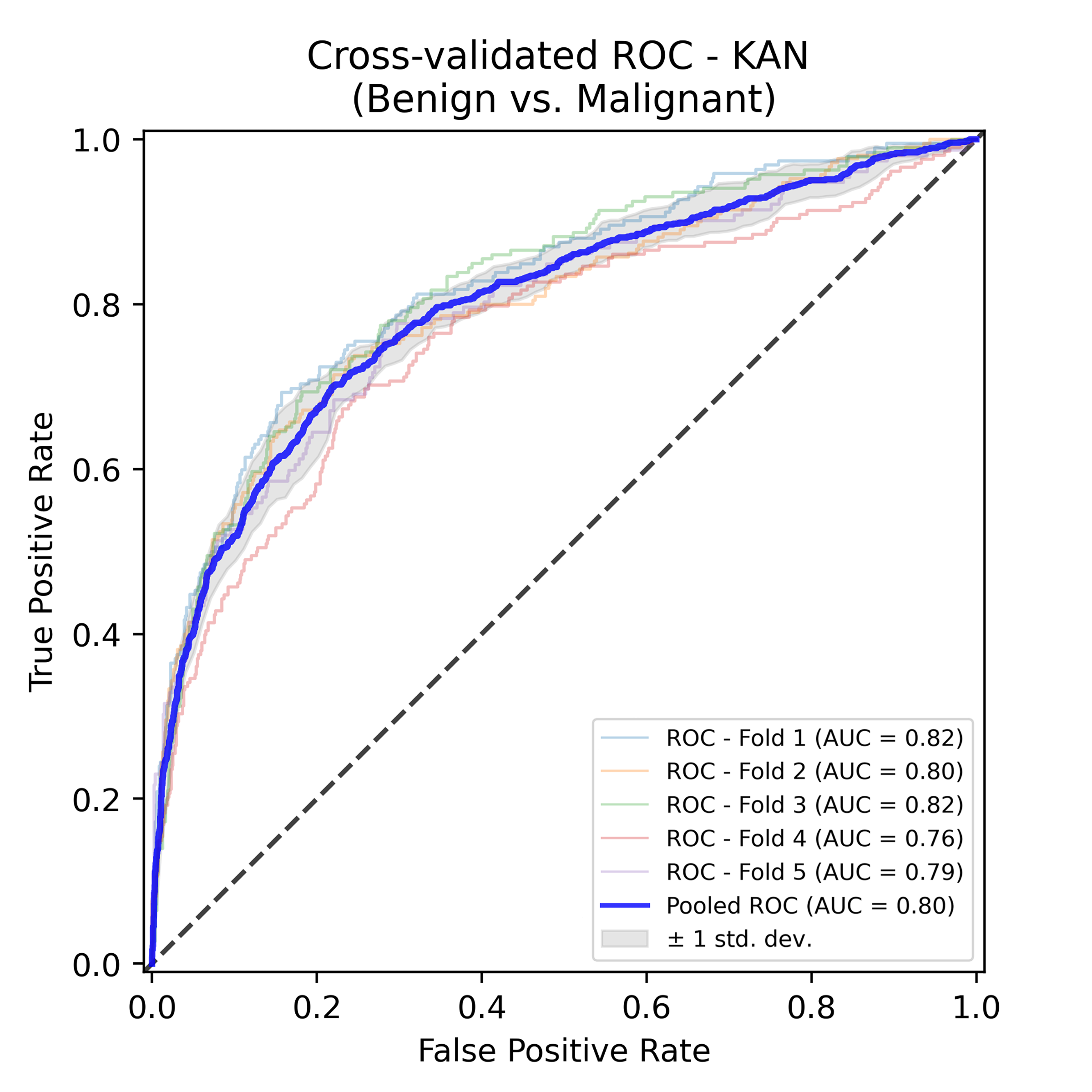}
        \caption{ROC-AUC curve of the KAN-based classifier.}
        \label{fig:roc_kan}
    \end{subfigure}
    \caption{ROC-AUC curves of the KAN and MST classifiers.}
    \label{fig:roc_comparison}
\end{figure}

The slice-wise transformer, fine-tuned with the KAN, achieved consistent and significant performance improvement across the five-fold cross-validations (Figure \ref{fig:roc_kan}). Compared to the standard MST baseline, our approach demonstrated a clear improvement in AUC getting AUC=0.8$\pm$0.02 while also showing enhanced robustness in handling class-imbalanced scenarios. These improvements highlight the effectiveness of combining self-supervised pretraining with the flexible KAN architecture. Our results demonstrate that this approach enables adaptation to a new task-detecting high probability of malignancy lesions with improved performance.

Our findings highlight that the integration of the MST with a KAN classifier yields superior performance in breast lesion classification compared to transformer-based approaches. The slice-wise approach provides a practical trade-off between computational efficiency and fine-grained spatial analysis, while attention-based saliency maps enhance interpretability by highlighting image regions that influence the model’s decision-making. Moreover, the adoption of KANs introduces a more flexible and interpretable feature space improving generalization across heterogeneous patient cohorts. 

However, this study has several limitations. First, the model was trained and evaluated on a relatively limited dataset, which may restrict its ability to generalize to broader clinical scenarios, particularly in the presence of rare lesion types or variations in imaging protocols. Additionally, while the slice-wise approach improves computational efficiency, it may not fully capture 3D contextual information found in volumetric breast imaging, potentially limiting the detection of subtle or spatially diffuse lesions. The current method also relies on expert-annotated labels, which may introduce subjectivity or inter-observer variability.

Future work will focus on expanding the dataset to include larger and more diverse patient cohorts, incorporating contextual data to capture relationships across sequential data slices. Furthermore, exploring self-supervised learning strategies could mitigate the dependency on large amounts of labeled data. We also aim to investigate domain adaptation techniques to enhance robustness across imaging devices and clinical settings.

\section{CONCLUSION}
We demonstrated the potential of adapting foundation model-based slice-wise transformers, such as MST, for robust lesion detection in breast DCE-MRI, particularly in high-risk breast cancer populations. We applied a transfer learning strategy and used MST embeddings with our KAN classifier, resulting in improved lesion classification. Attention maps enhance transparency by identifying the regions most relevant for lesion detection, providing valuable insights for clinical integration.   




\bibliography{report} 

\begin{thebibliography}{10}

\bibitem{10.1001/jamainternmed.2013.11958}
Stout, N.~K., Nekhlyudov, L., Li, L., Malin, E.~S., Ross-Degnan, D., Buist, D.
  S.~M., Rosenberg, M.~A., Alfisher, M., and Fletcher, S.~W., ``Rapid increase
  in breast magnetic resonance imaging use: Trends from 2000 to 2011,'' {\em
  JAMA Internal Medicine}~{\bf 174},  114--121 (01 2014).

\bibitem{10.1001/jamainternmed.2013.11963}
Wernli, K.~J., DeMartini, W.~B., Ichikawa, L., Lehman, C.~D., Onega, T.,
  Kerlikowske, K., Henderson, L.~M., Geller, B.~M., Hofmann, M., Yankaskas,
  B.~C., and for~the Breast Cancer Surveillance~Consortium, ``Patterns of
  breast magnetic resonance imaging use in community practice,'' {\em JAMA
  Internal Medicine}~{\bf 174},  125--132 (01 2014).

\bibitem{Lee2021-zz}
Lee, M.~V., Aharon, S., Kim, K., Sunn~Konstantinoff, K., Appleton, C.~M.,
  Stwalley, D., and Olsen, M.~A., ``Recent trends in screening breast {MRI},''
  {\em J Breast Imaging}~{\bf 4},  39--47 (Dec. 2021).

\bibitem{Killelea2013-fy}
Killelea, B.~K., Long, J.~B., Chagpar, A.~B., Ma, X., Soulos, P.~R., Ross,
  J.~S., and Gross, C.~P., ``Trends and clinical implications of preoperative
  breast {MRI} in medicare beneficiaries with breast cancer,'' {\em Breast
  Cancer Res Treat}~{\bf 141},  155--163 (Aug. 2013).

\bibitem{Strigel2017}
Strigel, R.~M., Burnside, E.~S., Elezaby, M., and et~al., ``Utility of bi-rads
  assessment category 4 subdivisions for screening breast mri,'' {\em AJR Am J
  Roentgenol}~{\bf 208}(6),  1392--1399 (2017).

\bibitem{Raza2008}
Raza, S., Chikarmane, S.~A., Neilsen, S.~S., Zorn, L.~M., and Birdwell, R.~L.,
  ``Bi-rads 3, 4, and 5 lesions: Value of us in management—follow-up and
  outcome,'' {\em Radiology}~{\bf 248}(3),  773--781 (2008).

\bibitem{Vlahiotis2018}
Vlahiotis, A., Griffin, B., Stavros, M. D.~F., and Margolis, J., ``Analysis of
  utilization patterns and associated costs of the breast imaging and
  diagnostic procedures after screening mammography,'' {\em ClinicoEconomics
  and Outcomes Research}~{\bf 10},  157--167 (2018).

\bibitem{Yu2022}
Yu, A.~C., Mohajer, B., and Eng, J., ``External validation of deep learning
  algorithms for radiologic diagnosis: A systematic review,'' {\em Radiology:
  Artificial Intelligence}~{\bf 4},  e210064 (May 2022).

\bibitem{wang2025foundation}
Wang, X. et~al., ``Foundation models in radiology: What, how, why, and why
  not,'' {\em Radiology}~{\bf 308},  e240597 (2025).

\bibitem{Jang2022M3T}
Jang, J., Hwang, D., and et~al., ``M3t: Three-dimensional medical image
  classifier using multi-plane and multi-slice transformer,'' in [{\em
  Proceedings of the 2022 IEEE/CVF Conference on Computer Vision and Pattern
  Recognition (CVPR)}{\nolinebreak\hspace{0.1em}]},  IEEE, New Orleans, LA, USA
  (June 2022).

\bibitem{Tustison2010}
Tustison, N.~J., Avants, B.~B., Cook, P.~A., Zheng, Y., Egan, A., Yushkevich,
  P.~A., and Gee, J.~C., ``{N4ITK}: Improved {N3} bias correction,'' {\em IEEE
  Transactions on Medical Imaging}~{\bf 29},  1310--1320 (June 2010).

\bibitem{10.1145/3383812.3383830}
Javadi, S., Dahl, M., and Pettersson, M.~I., ``Adjustable contrast enhancement
  using fast piecewise linear histogram equalization,'' in [{\em Proceedings of
  the 2020 3rd International Conference on Image and Graphics
  Processing}{\nolinebreak\hspace{0.1em}]},  {\em ICIGP '20},  57–61,
  Association for Computing Machinery, New York, NY, USA (2020).

\bibitem{Muller-Franzes2025}
Müller-Franzes, G., Khader, F., Siepmann, R., and et~al., ``Medical slice
  transformer for improved diagnosis and explainability on 3d medical images
  with dinov2,'' {\em Scientific Reports}~{\bf 15},  23979 (2025).
\newblock Received 11 February 2025; Accepted 25 June 2025; Published 04 July
  2025.

\bibitem{oquab2023dinov2}
Oquab, M., Darcet, T., Moutakanni, T., Vo, H.~V., Szafraniec, M., Khalidov, V.,
  Fernandez, P., Haziza, D., Massa, F., El-Nouby, A., Howes, R., Huang, P.-Y.,
  Xu, H., Sharma, V., Li, S.-W., Galuba, W., Rabbat, M., Assran, M., Ballas,
  N., Synnaeve, G., Misra, I., Jegou, H., Mairal, J., Labatut, P., Joulin, A.,
  and Bojanowski, P., ``Dinov2: Learning robust visual features without
  supervision,'' (2023).

\bibitem{saha2021dce}
Saha, A., Harowicz, M.~R., Grimm, L.~J., Weng, J., Cain, E.~H., Kim, C.~E.,
  Ghate, S.~V., Walsh, R., and Mazurowski, M.~A., ``Dynamic contrast-enhanced
  magnetic resonance images of breast cancer patients with tumor locations.''
  \url{https://doi.org/10.7937/TCIA.e3sv-re93} (2021).
\newblock The Cancer Imaging Archive [Data set].

\bibitem{liu2024kan}
Liu, Z., Wang, Y., Vaidya, S., Ruehle, F., Halverson, J., Solja{\v{c}}i{\'c},
  M., Hou, T.~Y., and Tegmark, M., ``Kan: Kolmogorov-arnold networks,'' {\em
  arXiv preprint arXiv:2404.19756}  (2024).

\bibitem{10.1007/11538059_91}
Han, H., Wang, W.-Y., and Mao, B.-H., ``Borderline-smote: a new over-sampling
  method in imbalanced data sets learning,'' in [{\em Proceedings of the 2005
  International Conference on Advances in Intelligent Computing - Volume Part
  I}{\nolinebreak\hspace{0.1em}]},  {\em ICIC'05},  878–887, Springer-Verlag,
  Berlin, Heidelberg (2005).

\bibitem{lin2018focallossdenseobject}
Lin, T.-Y., Goyal, P., Girshick, R., He, K., and Dollár, P., ``Focal loss for
  dense object detection,'' (2018).

\end{thebibliography}
\bibliographystyle{spiebib} 

\end{document}